\documentclass[journal]{IEEEtran}

\ifCLASSINFOpdf
\else
   \usepackage[dvips]{graphicx}
\fi
\usepackage{url}

\usepackage[rgb]{xcolor}
\usepackage{hyperref}
\hypersetup{
    colorlinks=true,
    citecolor=blue,
}
\usepackage{colortbl}

\usepackage{graphicx}
\usepackage{cite}
\usepackage{amsmath}
\usepackage{booktabs}
\usepackage{multirow}

\begin{document}

\title{Multi-speaker Text-to-speech Training with Speaker Anonymized Data}

\author{Wen-Chin Huang, \IEEEmembership{Member, IEEE}, Yi-Chiao Wu, and Tomoki Toda \IEEEmembership{Member, IEEE}
\thanks{This work was partly supported by JST CREST under Grant Number JPMJCR19A3. Wen-Chin Huang is with the Graduate School of Informatics, Nagoya University, Japan (e-mail: wen.chinhuang@g.sp.m.is.nagoya-u.ac.jp). Yi-Chiao Wu is with Meta, USA. Tomoki Toda is with the Information Technology Center, Nagoya University, Japan.}
\thanks{Audio samples are available at: \url{https://unilight.github.io/Publication-Demos/publications/sa-tts-spl/index.html}}
}

\markboth{Journal of \LaTeX\ Class Files, Vol. 14, No. 8, August 2015}
{Shell \MakeLowercase{\textit{et al.}}: Bare Demo of IEEEtran.cls for IEEE Journals}
\maketitle

\begin{abstract}
The trend of scaling up speech generation models poses a threat of biometric information leakage of the identities of the voices in the training data, raising privacy and security concerns.
In this paper, we investigate training multi-speaker text-to-speech (TTS) models using data that underwent speaker anonymization (SA), a process that tends to hide the speaker identity of the input speech while maintaining other attributes.
Two signal processing-based and three deep neural network-based SA methods were used to anonymize VCTK, a multi-speaker TTS dataset, which is further used to train an end-to-end TTS model, VITS, to perform unseen speaker TTS during the testing phase.
We conducted extensive objective and subjective experiments to evaluate the anonymized training data, as well as the performance of the downstream TTS model trained using those data.
Importantly, we found that UTMOS, a data-driven subjective rating predictor model, and GVD, a metric that measures the gain of voice distinctiveness, are good indicators of the downstream TTS performance.
We summarize insights in the hope of helping future researchers determine the goodness of the SA system for multi-speaker TTS training.
\end{abstract}


\IEEEpeerreviewmaketitle

\section{Introduction}
\label{sec:intro}



\IEEEPARstart{S}{caling} up speech generation models in terms of both model size and training data has become a trend in the research community. For instance, multiple works have reported training their model on more than 100k hours of data \cite{basetts, naturalspeech3, audiobox}. However, the larger the model becomes, the more likely it \textit{memorizes} parts of the training data \cite{memorization-in-llm}. In the task of speech generation, memorization results in biometric information leakage, which causes security and privacy issues. For instance, a speaker whose voice was used in the training data may be memorized and maliciously generated to be used to spoof a voice authentication system. With the increasing interest in data privacy protection, including legal movements like the European General Data Protection Regulation (GDPR), scaling up speech generation models will become difficult.

A possible solution to the above-mentioned problem is to train speech generation models with data that underwent a so-called speaker anonymization (SA) process, which attempts to erase the biometric information of the input speech while preserving certain properties. 
This research field has been greatly advanced and promoted by the voice privacy challenge (VPC) series \cite{vpc2020, vpc22-plan}, where the organizers established a series of standardized dataset settings and evaluation protocols.
In VPC, the two primary metrics adopted to evaluate SA systems were the equal error rate (\textbf{EER}) calculated with an automatic speaker verification model, and the word error rate (\textbf{WER}) obtained from an automatic speech recognition model. The former was called the \textit{privacy} and the latter the \textit{utility} metric. On the other hand, evaluating these SA systems in the context of speech generation model training has not yet been investigated, and it is unknown whether an SA system that performs well in terms of EER and WER can also excel in the downstream speech generation task.

In this paper, as a proof-of-concept, we investigate training a multi-speaker text-to-speech (TTS) model with speaker anonymized data, with the hope of providing a reasonable framework to evaluate SA systems in terms of the performance of the downstream TTS task. We adopted two signal processing-based and three deep neural network-based SA methods and used them to anonymize a multi-speaker TTS dataset. These anonymized datasets are further used to train multi-speaker TTS models, which are then evaluated on the task of unseen speaker TTS. An extensive experimental evaluation was conducted, as we reported objective metrics from VPC'22, as well as subjective evaluation results on the anonymized training data and the downstream TTS output.
The contributions of this work are as follows:
\begin{itemize}
    \item This is the first work to investigate the impact of speaker-anonymized training data on a downstream speech generation task, specifically, multi-speaker TTS training.
    \item We identified the relationship between the performance measurements of SA systems and the downstream multi-speaker models and provided guidelines for future researchers to develop better SA systems.
\end{itemize}

\section{Problem Formulation}
\label{sec:formulation}

\begin{figure}[t]
	\centering
	\includegraphics[width=0.9\linewidth]{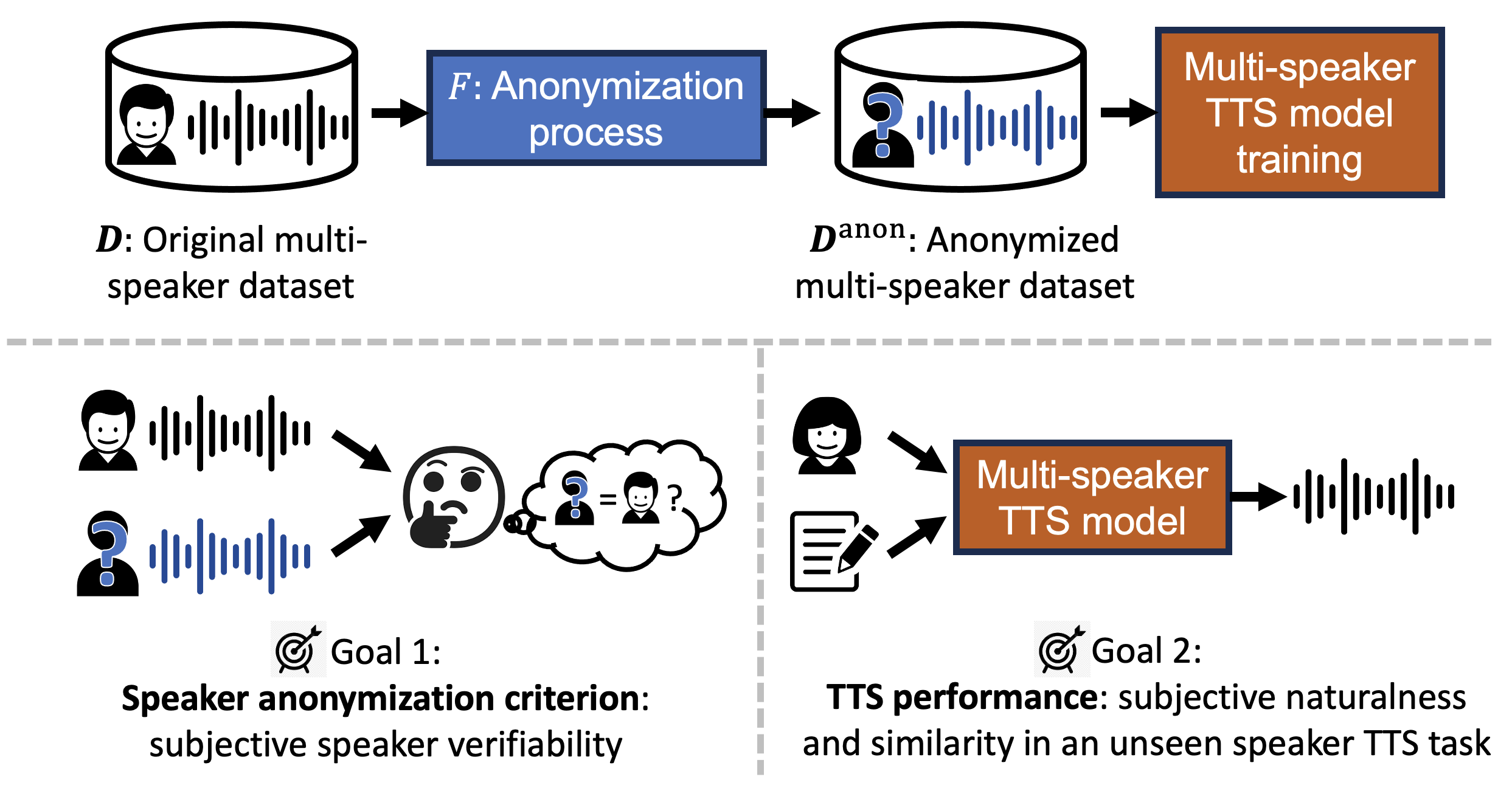}
         \centering
	\caption{\label{fig:overview}Problem formulation and goals of this work.}	
        \vspace{-10pt}
\end{figure}

The problem formulation and goals are illustrated in Figure~\ref{fig:overview}. Suppose we have an initial user dataset that we wish to erase biometric information from, denoted as $\textbf{D}$. An anonymization process is performed on $\textbf{D}$ to obtain the anonymized version, denoted as $\textbf{D}^{\text{anon}}=F(\textbf{D})$. It is then used to train a multi-speaker TTS model.
There are two goals in this problem setting. First, $\textbf{D}^{\text{anon}}$ needs to fulfill a pre-defined \textbf{speaker anonymization criterion}.
Second, the \textbf{TTS performance} should be maximized, which is evaluated on a pre-defined downstream task.
For the anonymization criterion, in addition to adopting the objective privacy metric (i.e., EER) as in the VPCs, we also measure the subjective speaker verifiability as proposed in VPC \cite{vpc2020}.
For the downstream task, we consider unseen speaker TTS, which is to generate the voice of the designated speaker given the text and a short reference speech sample. Other speech generation tasks, such as speaker adaptive TTS and speech enhancement, will be left as future work.

As we will describe in Section~\ref{sec:sa-systems}, the anonymization process can include a data-driven model trained on some other speech datasets.
Here we assume that the training data of the SA model does not need to be anonymized (based on the condition that, for instance, they are public and we are authorized to use them). We consider this setting to be practical because, as we will show in later sections, the amount of training data for most SA systems is much smaller than those used to train large speech generation models.
We would like to once again emphasize that the focus of this paper is on properly using first-party data while protecting users' privacy. 

\section{Speaker Anonymization Systems}
\label{sec:sa-systems}

In this section, we describe the speaker anonymization systems adopted in this work. These systems were adopted either because they are representative (for instance, baselines of VPC'22), or because they are easy to reproduce (for instance, with open-source implementations).

\subsection{Signal processing based systems}

Signal processing-based SA systems enjoy the benefit of being free from training and efficient inference. In this work, we adopt two methods, each of which modifies a certain speech parameter obtained from speech analysis.

\subsubsection{Pitch shift}

Simply shifting the pitch of the input speech was shown to be on par with the baselines in VPC'22 \cite{vpc22_pitch_shift}, which was implemented using a time-scaling approach. However, in our preliminary experiments, we found it resulted in poor perceptual quality that hurt the performance of the downstream TTS model.
We thus adopted a python wrapper \cite{pyworld} of WORLD, a high-quality vocoder \cite{world}. Anonymization is done by randomly shifting the extracted f0 sequence up or down within 3 to 5 semitones and then synthesizing the anonymized waveform with the other speech parameters.

\subsubsection{VPC'22 B2: Spectral envelope modification}

In contrast to modifying the frequency component, the B2 baseline in VPC'22 modified the spectral envelope of the input speech, resulting in a change in the timbre. It is based on the idea proposed in \cite{vpc22b2}. First, linear predictive coding (LPC) coefficients extracted from the input speech are converted to pole positions. Then, the phase of poles with non-zero imaginary parts is raised to the power of the McAdams’ coefficient $\alpha$ such that transformed poles have new, shifted phases of $\phi^{\alpha}$. The poles are then converted back to LPC coefficients and thus the anonymized waveform. The VPC'22 B2 version sampled the McAdams’ coefficient from a uniform distribution as $\alpha \sim \mathcal{U}(0.5, 0.9)$. We adopted the implementation from the VoicePAT toolkit \cite{voicepat}.

\begin{figure}[t]
	\centering
	\includegraphics[width=\linewidth]{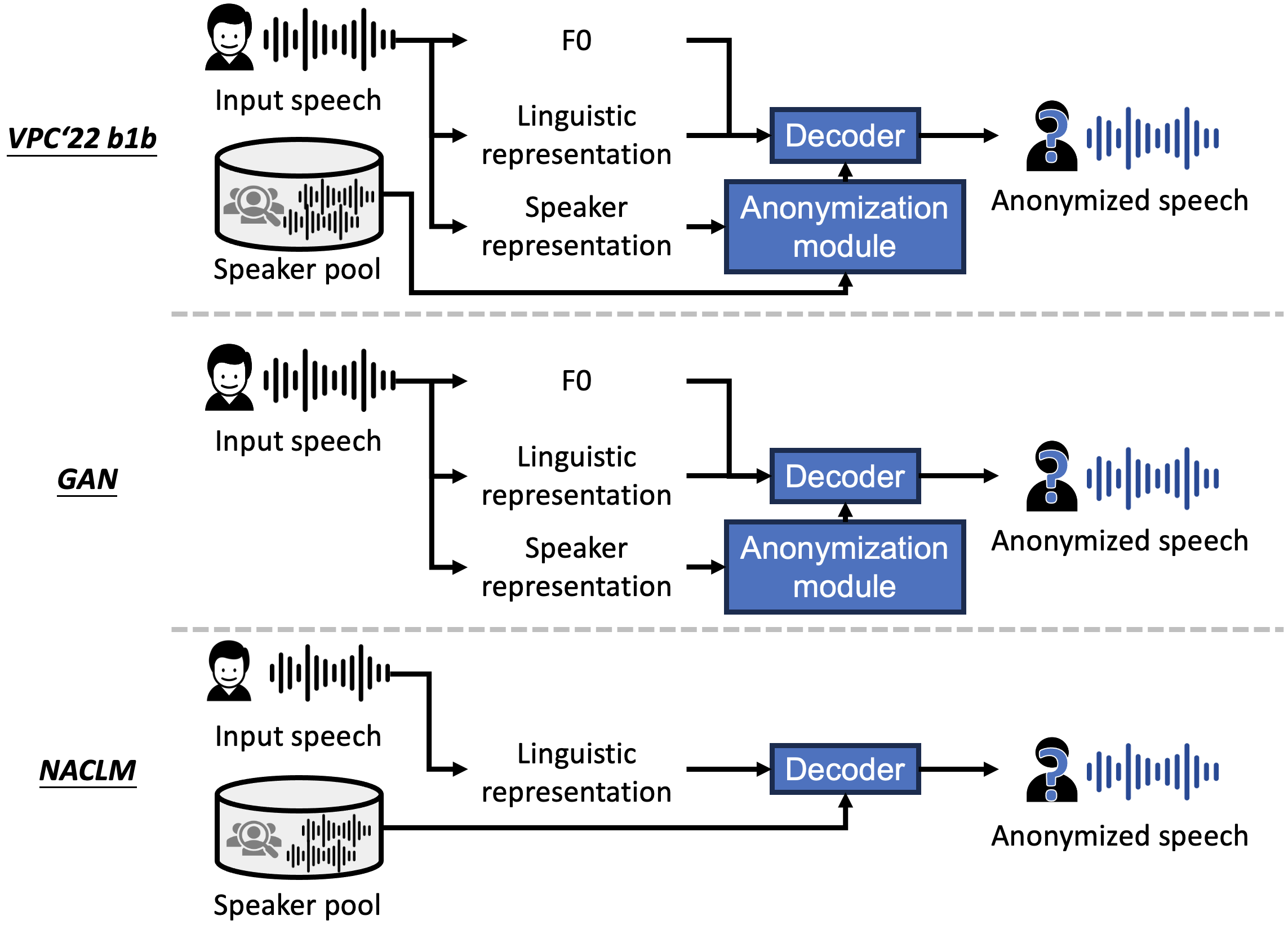}
         \centering
	\caption{\label{fig:dnn-based}Deep neural network-based speaker anonymization systems.}	
        \vspace{-10pt}
\end{figure}

\subsection{Deep neural network based systems}

Compared to signal processing-based methods, deep neural network (DNN)-based SA systems were reported to perform better in almost all metrics in VPC'22 \cite{vpc22-results}. In this work, we adopt three DNN-based systems, as illustrated in Figure~\ref{fig:dnn-based}. All DNN-based methods attempt to factorize the input speech into several components and change the speaker representation to achieve anonymization.

\subsubsection{VPC'22 B1b}

The VPC'22 B1b baseline is an improved version of the system proposed in \cite{fang19_ssw}. It first factorizes the input speech into f0, linguistic representation, and speaker representation. The linguistic representation is the frame-level output of the encoder of an automatic speech recognition (ASR) model, and the speaker representation is the x-vector \cite{x-vector}. The anonymization module finds 200 speaker representations in a pre-defined speaker pool that are farthest from the input speaker representation and then averages random 100 of them to form the anonymized speaker representation. The decoder takes the f0, linguistic representation and the anonymized speaker representation to generate the final anonymized speech.
We followed the official implementation released by the VPC'22 organizers \cite{vpc22baseline}

\subsubsection{GAN}

The GAN method was proposed in \cite{gan-based} and differs from VPC'22 B1b in two aspects. First, the linguistic representation is the phoneme sequence from the ASR model instead of the frame-level output, which in practice contains fewer speaker information, leading to better anonymization performance. Second, instead of relying on a speaker pool, the anonymization module is now a generative adversarial network that was trained to map a normal distribution to the approximate distribution of the x-vector. The sampled x-vector is ensured to have a cosine similarity smaller than 0.7 with the input speaker x-vector. We adopted the implementation from the VoicePAT toolkit \cite{voicepat}.

\subsubsection{NACLM}

The final method is based on neural audio codec language models (NACLMs), originated from AudioLM \cite{audiolm}.
We include this system for its good EER performance, as we will show in Section~\ref{ssec:exp-evaluation-sa-training-data}.
The idea is to condition a language model with the so-called \textit{semantic tokens} and an \textit{acoustic prompt} to generate acoustic tokens (also known as neural codec), which have the same content as the semantic tokens and the same acoustic condition as the acoustic tokens. They are then passed to a synthesizer to obtain the final waveform. Here, the linguistic representation, which is the HuBERT \cite{hubert} sequence of the input speech, is the semantic tokens, and the acoustic prompt is an acoustic token sequence randomly sampled from the speaker pool.
We used the official implementation provided by the authors, which was based on Bark, an open-source NACLM-based TTS system \cite{naclm-github}.

\section{Experimental Evaluation}

There are two parts of the evaluation: in the first part (Section~\ref{ssec:exp-evaluation-sa-training-data}), we evaluated the anonymized training data, and in the second part (Section~\ref{ssec:exp-evaluation-tts}), we evaluated the TTS systems trained on the anonymized data. Finally, in Section~\ref{ssec:exp-indicators} we discuss the relationship between the metrics in the first and second parts.

\subsection{Data and implementation}

The dataset used to train and evaluate the TTS systems (i.e., $\textbf{D}$ in Section~\ref{sec:formulation}) was VCTK \cite{vctk}. All samples were downsampled to 16kHz. Following \cite{yourtts}, we excluded 11 speakers for evaluation. During inference, the input to the TTS system included the text of the last 50 samples of each evaluation speaker, and the \texttt{005} sample as the reference. Following \cite{yourtts}, the TTS system was pre-trained on LJSpeech \cite{ljspeech}. The TTS system was VITS \cite{vits} with x-vectors \cite{x-vector} as the speaker embedding. We used the implementation provided in ESPnet2-TTS \cite{espnet2-tts}. 

The implementation of the SA systems was described in Section~\ref{sec:sa-systems}.
For the training data of the DNN-based SA systems, both VPC'22 B1b and GAN shared a similar setting to that of VPC'22: the ASR model, x-vector extractor, and decoder were trained on LibriSpeech \cite{librispeech}, VoxCeleb 1 \& 2 \cite{voxceleb1, voxceleb2}, LibriTTS train-clean-100 \cite{libritts}, respectively. The speaker pool was the LibriTTS train-other-500 set. For the NACLM system, since Bark was directly used without any re-training, the setting was largely different from VPC'22 B1b and GAN, resulting in an unfair comparison. Readers should note this difference. Finally, the LibriSpeech test set was used to evaluate the SA systems, as in VPC.

\subsection{Evaluation Metrics and Protocols}

The objective evaluation of the SA systems was carried out with the VoicePAT toolkit \cite{voicepat}.
As in the VPC series, the \textbf{EER} (the larger the better) and the \textbf{WER} (the lower the better) were reported.
In addition, we also reported the gain of voice distinctiveness (\textbf{GVD}) proposed in \cite{vpc2020}. The larger the GVD, the better the distinctiveness is maintained throughout the anonymization process. Finally, we added the \textbf{UTMOS} score, a widely used perceptual rating predictor trained with human ratings \cite{utmos}, which is the larger the better.

As the ultimate goal of this work is to train TTS systems with high quality, it is essential to conduct listening tests to assess the perceptual quality. In this work, we assessed the subjective \textit{naturalness} and speaker \textit{similarity}.
Listeners were asked to evaluate the naturalness of the speech on a 5-point scale, and the numbers the higher the better.
For similarity, following the protocol in the voice conversion challenge 2020 \cite{vcc2020}, a natural speech from the reference speaker and a generated speech were presented, and listeners were asked to judge whether the two samples were produced by the same speaker on a 4-point scale.
We evaluate naturalness and similarity for both SA and TTS, and they are referred to as \textbf{SA-NAT}, \textbf{SA-SIM}, \textbf{TTS-NAT} and \textbf{TTS-SIM}, respectively.
Notably, for SA-SIM, the goal was to assess how \textit{dissimilar} the anonymized speech was to the original speech, thus the lower the better. In contrast, for TTS-SIM, the similarity score was the higher the better.
We used crowd-sourcing to recruit 500 listeners and obtained 2750 ratings per system.
Recordings of the natural samples (GT) were also included to serve as the upper bound.

\begin{table*}[t]
\centering
\caption{Evaluation results of the speaker anonymized data and the downstream TTS model. Bold face indicates the best performance among the five SA systems.}
\label{tab:results}
\begin{tabular}{@{}c|cccccc|cc@{}}
\toprule
\multirow{2}{*}{System} & \multicolumn{6}{c|}{SA evaluation}                                                                                          & \multicolumn{2}{c}{TTS evaluation}                  \\ \cmidrule(l){2-9} 
                        & EER $\uparrow$ & WER $\downarrow$ & GVD $\uparrow$ & UTMOS $\uparrow$ & SA-NAT $\uparrow$        & SA-SIM $\downarrow$      & TTS-NAT $\uparrow$       & TTS-SIM $\uparrow$       \\ \midrule
Natural                 & --             & 2.98             & --             & 3.96             & 3.95 $\pm$ 0.03          & 3.62 $\pm$ 0.03          & 3.84 $\pm$ 0.03          & 3.66 $\pm$ 0.02          \\
Unanonymized            & --             & --               & --             & --               & --                       & --                       & 3.82 $\pm$ 0.03          & 2.59 $\pm$ 0.04          \\ \midrule
Pitch shift             & 4.94           & \textbf{3.18}    & \textbf{-0.66} & 3.31             & 2.92 $\pm$ 0.04          & 2.39 $\pm$ 0.04          & 3.13 $\pm$ 0.03          & \textbf{2.41 $\pm$ 0.04} \\
VPC'22 B2               & 5.91           & 12.02            & -2.53          & 2.21             & 1.48 $\pm$ 0.03          & 1.99 $\pm$ 0.03          & 1.82 $\pm$ 0.03          & 1.82 $\pm$ 0.03          \\ \midrule
VPC'22 B1b              & 9.28           & 3.97             & -8.85          & \textbf{3.90}    & \textbf{3.58 $\pm$ 0.03} & 2.06 $\pm$ 0.02          & 3.04 $\pm$ 0.03          & 1.61 $\pm$ 0.03          \\
GAN                     & 39.67          & 7.87             & -3.70          & 3.73             & 3.34 $\pm$ 0.04          & \textbf{1.28 $\pm$ 0.02} & \textbf{3.42 $\pm$ 0.03} & 1.85 $\pm$ 0.04          \\
NACLM                   & \textbf{45.77} & 7.12             & -2.40          & 3.45             & 2.84 $\pm$ 0.04          & 1.36 $\pm$ 0.02          & 2.53 $\pm$ 0.03          & 2.02 $\pm$ 0.04          \\ \bottomrule
\end{tabular}
\vspace{-10pt}
\end{table*}

\begin{figure}[t]
	\centering
	\includegraphics[width=\linewidth]{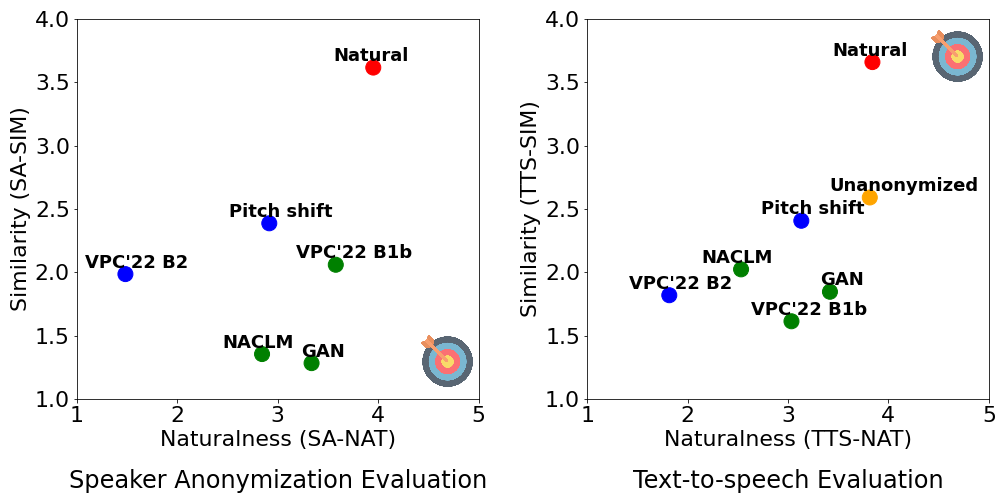}
         \centering
	\caption{\label{fig:scatter_scores}Scatter plots of the SA and TTS subjective evaluation results. The goal icons are located at the ideal score positions. Blue and green dots indicate signal processing-
 based and deep neural network-based systems, respectively.}
        \vspace{-5pt}
\end{figure}



\subsection{Evaluation results of the anonymized training data}
\label{ssec:exp-evaluation-sa-training-data}

We first look at the SA evaluation results, as shown in Table~\ref{tab:results}.
We observed that no single system was dominant in all six metrics. As perceptual evaluation has been overlooked in the SA literature, we are especially interested in the subjective results, which were visualized in Figure~\ref{fig:scatter_scores}.
As an ideal SA system should give a high SA-NAT score and a low SA-SIM score, the results showed that no single system was dominant in both SA-NAT and SA-SIM. Specifically, VPC'22 B1b and GAN had the best SA-NAT and SA-SIM scores, respectively. We also found that DNN-based SA systems are better than signal-processing-based systems.

We are also interested in whether primary objective metrics adopted by VPC (namely, EER and WER) correlate well with subjective metrics (SA-NAT and SA-SIM). The linear correlation coefficients between EER and SA-SIM, WER and SA-NAT are -0.946 and -0.828, respectively. While EER correlates with SA-SIM, the correlation between WER and SA-NAT is lower. We further found that the linear correlation coefficient between UTMOS and SA-NAT is 0.984. This suggests that compared to WER, UTMOS is a better indicator of subjective naturalness.

\vspace{-5pt}
\subsection{Evaluation results of the downstream TTS task}
\label{ssec:exp-evaluation-tts}

We then look at the TTS evaluation results, which are shown in Table~\ref{tab:results} and visualized in Figure~\ref{fig:scatter_scores}.
The \textit{unanonymized} system refers to a TTS model trained with unanonymized data directly, thus serving as an upper bound.
As an ideal TTS system should yield high TTS-NAT and TTS-SIM scores, there was again no single system dominant in both TTS-NAT and TTS-SIM, with GAN and pitch shift being the best systems in terms of TTS-NAT and TTS-SIM, respectively.

\begin{table}[t]
\centering
\caption{Linear correlation coefficients between the SA evaluation metrics and the TTS evaluation results. Bold indicates a strong correlation ($>0.7$).}
\label{tab:correlation}
\begin{tabular}{@{}ccc|cc@{}}
\toprule
                                                                         &                       &        & \multicolumn{2}{c}{TTS evaluation} \\ \cmidrule(l){4-5} 
                                                                         &                       &        & TTS-NAT          & TTS-SIM         \\ \midrule
\multirow{6}{*}{\begin{tabular}[c]{@{}c@{}}SA\\ evaluation\end{tabular}} & \multirow{4}{*}{Obj.} & WER    & \textbf{-0.785}  & -0.529          \\
                                                                         &                       & EER    & 0.231            & -0.061          \\
                                                                         &                       & GVD    & -0.220           & \textbf{0.827}  \\
                                                                         &                       & UTMOS  & \textbf{0.874}   & 0.341           \\ \cmidrule(l){2-5} 
                                                                         & \multirow{2}{*}{Sub.} & SA-NAT & \textbf{0.929}   & 0.469           \\
                                                                         &                       & SA-SIM & 0.477            & \textbf{0.864}  \\ \bottomrule
\end{tabular}
\vspace{-10pt}
\end{table}

\vspace{-5pt}
\subsection{Important indicators of the TTS performance}
\label{ssec:exp-indicators}

In this subsection, we investigate whether we can \textit{determine the goodness of an SA method before actually training and evaluating the downstream TTS system}.
From Table~\ref{tab:results}, we calculated the linear correlation coefficient of each metric on the SA evaluation side with each of the TTS evaluation results and summarized in Table~\ref{tab:correlation}. Although the best indicators for TTS-NAT and TTS-SIM were SA-NAT and SA-SIM, respectively, conducting listening tests is costly and should be avoided. We then seek to find important indicators using only objective metrics.

For TTS-NAT, UTMOS has the highest correlation (0.874), with WER being the second highest (0.785). This again shows the suitability of using UTMOS as the utility metric in the context of SA. As for TTS-SIM, GVD was the only metric that provided a strong correlation (0.827). We would like to highlight this important result: although VPC'22 used it only as a secondary metric the GVD metric measures the SA system's ability to preserve voice distinctiveness (i.e., diversity), which is especially important for multi-speaker TTS training. One may consider an extreme case where the SA system maps all speakers in the input dataset to a canonical speaker. Such an SA system can give us a high EER, but possibly a GVD approaching minus infinity. The resulting anonymized dataset essentially becomes a single-speaker dataset, which would not be feasible for multi-speaker training.

\section{Conclusion and Future Directions}

In this work, we trained multi-speaker TTS models using speaker anonymized data. Five SA systems were used to anonymize the training data of an end-to-end TTS model, which is further evaluated on an unseen TTS task. From our extensive experimental results, we conclude that a good SA system for anonymizing training data for multi-speaker TTS should ensure \textbf{(1) a high UTMOS score indicating high-quality output and (2) a high GVD indicating low loss of speech diversity}. Future works include improving current SA systems towards the above-mentioned metrics and exploring more speech generation tasks.


Finally, we would like to raise attention to a fundamental question: \textbf{What is a valid speaker anonymization threshold?}
Readers should note that SA-SIM and TTS-SIM have different meanings: \textbf{SA-SIM is the anonymization threshold to be satisfied, and TTS-SIM is a metric for the downstream TTS task to be maximized}.
Only when the SA-SIM (or EER) threshold is met does maximizing the TTS metrics become meaningful.
For instance, looking at Table~\ref{tab:results}, if we set the threshold to be an SA-SIM score of 1.5, then only GAN and NACLM are qualified SA systems; if we set the threshold to be 2.5, then all systems are qualified.
The determination of the threshold should come prior to the subsequent SA model development.
With that being said, we believe this is a question out of the scope of computer science research. requiring discussion between researchers, legal specialists, or even the public to reach a consensus.

\bibliographystyle{IEEEtran}
\bibliography{ref}

\end{document}